\documentclass[prodmode, acmtap]{ci2018}

\acmVolume{2}
\acmNumber{3}
\acmArticle{1}
\articleSeq{1}
\acmYear{2010}
\acmMonth{5}

\usepackage{latexsym}
\usepackage{graphicx}
\usepackage{mathptmx}
\usepackage{float}
\usepackage[normalem]{ulem}
\usepackage{wrapfig}
\usepackage[autostyle]{csquotes}

\usepackage[ruled]{algorithm2e}
\SetAlFnt{\algofont}
\SetAlCapFnt{\algofont}
\SetAlCapNameFnt{\algofont}
\SetAlCapHSkip{0pt}
\IncMargin{-\parindent}

\title{Simon's Anthill: Mapping and Navigating Belief Spaces}
\author{\MakeUppercase{Philip Feldman and Aaron Dant} \affil{ASRC Federal}
\MakeUppercase{WAYNE LUTTERS}
\affil{University of Maryland Baltimore County}
}

\begin{abstract}
See "abstract" subsection below
\end{abstract}

\begin{document}

\maketitle

\begin{figure}[h]
	\centering
	\begin{minipage}{.5\textwidth}
		\centering
		\fbox{\includegraphics[height=18em]{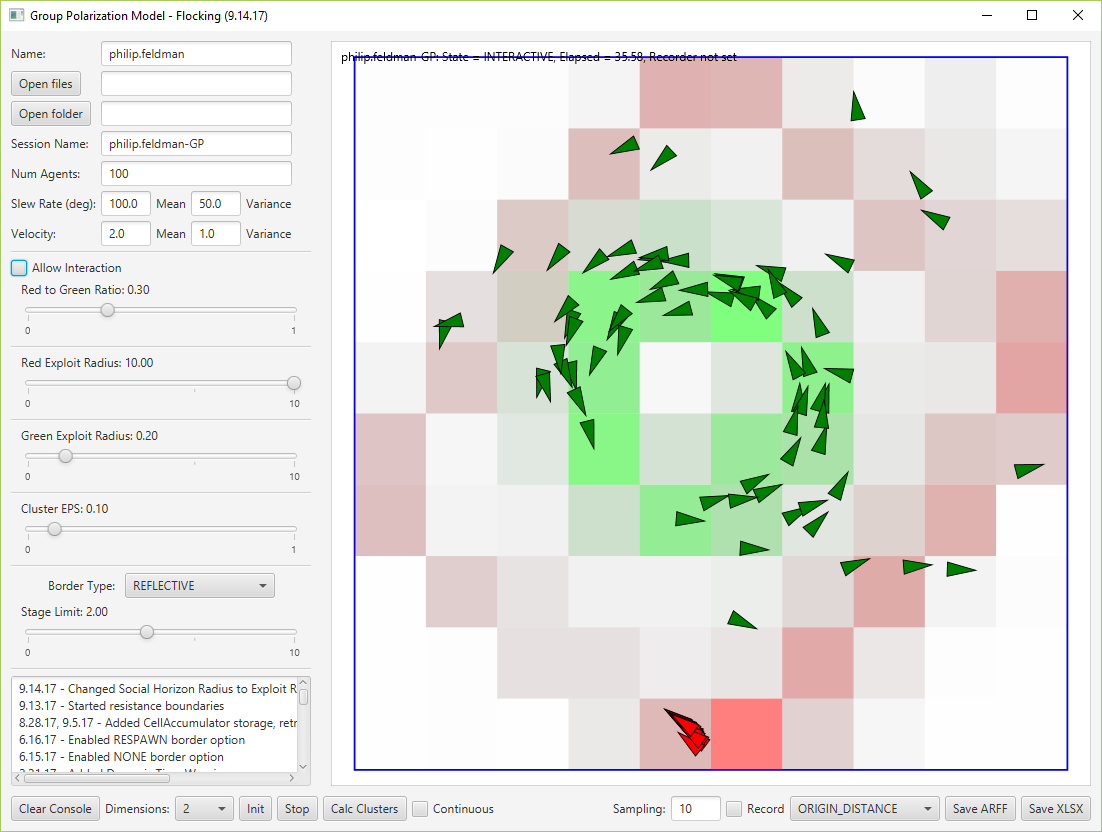}}
		\caption{\label{fig:screenshot}Flocking (green) and stampeding (red) agents}
	\end{minipage}%
	\begin{minipage}{.5\textwidth}
		\centering
		\fbox{\includegraphics[height=18em]{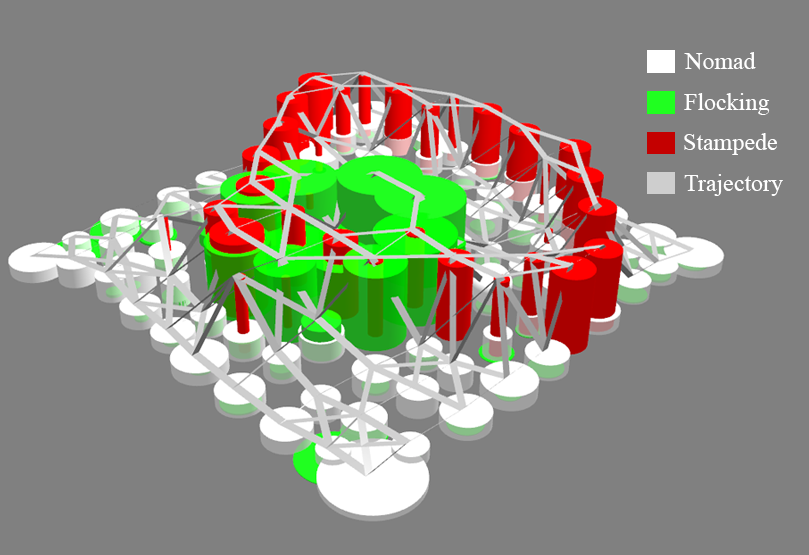}}
		\caption{\label{fig:anthill}Belief Map, using simulated lexical trajectories}
	\end{minipage}%
\end{figure}

\section{\MakeUppercase{Introduction}}

In the parable of Simon's Ant \cite{simon1996sciences}, an ant follows a complex path along a beach on to reach its goal. The story shows how the interaction of simple rules and a complex environment result in complex behavior. But this relationship can be looked at in another way -- given path and rules, we can infer the environment. With a large population of agents -- human or animal -- it should be possible to build a detailed map of a population's social and physical environment. In this abstract, we describe the development of a framework to create such \enquote{maps}  of human belief space. These maps are built from the combined trajectories of a large number of agents. Currently, these maps are built using multidimensional agent-based simulation, but the framework is designed to work using data from computer-mediated human communication. Maps incorporating human data should support visualization and navigation of the \enquote{plains of research} , \enquote{fashionable foothills}  and \enquote{conspiracy cliffs}  of human belief spaces.

\section {\MakeUppercase{Previous Work}}
Coming to consensus is rarely an act of compromise. Imagine the passengers of a car, lost and stopped at an intersection with no GPS. Each passenger has a different idea of where to go. If everyone compromises, then the car will stay at the intersection. Instead, the passengers need to find a way to agree on a single direction. \cite{moscovici1994conflict}  studied how these kinds of groups first \textit{simplify} a complex problem, and then align themselves for or against this simplification. A neurological basis for this sense of \enquote{shared direction} has been uncovered by  \cite{Stephens14425}, who showed that fMRI patterns in the brains of storytellers and listeners aligned as a function of shared understanding.

Finding consensus has similarities to collective animal behavior in physical space. Flocking has been shown to represent a form of group cognition \cite{petit2010decision}. Schooling fish are better at sensing food in noisy environments \cite{grunbaum1998schooling}. Danchen et. al. showed that animals and humans both use inadvertent social information to influence decisions about environmental quality and appropriateness \cite{danchin2004public}. Agent-based simulation has proven to be a particularly effective mechanism for modeling these animal patterns \cite{Reynolds:1987:FHS:37401.37406}, while also proving effective at modeling cognitive actions such as culture dissemination \cite{sen2013sociophysics}. The common properties of these and other collective behaviors have been explored by \cite{Olfati-Saber2007}, who provides a theoretical framework for analysis of consensus algorithms for multi-agent networked systems. 

\section{\MakeUppercase{The Model}}
Our model is based on two ideas: 1) that human navigation through \textit{belief space} (a subset of information space that contains items associated with opinions) is analogous to animal motion through physical space, and 2) that the \textit{digital inadvertent social information} provided by humans interacting with the belief environment can be used to characterize the underlying belief space. To explore these concepts in depth, we built a standalone simulator (Fig.\ref{fig:screenshot}), based on the Reynolds model \cite{Reynolds:1987:FHS:37401.37406}, that represents belief space as a hypercube composed of labeled cells, that supports manipulating the following elements:

\begin{enumerate}
	\item \textit{Dimension} -- Since the number of beliefs that a person may hold is not limited by physical space, the Reynolds algorithm was modified to work with arbitrary numbers of dimensions.
	\item \textit{Velocity} -- Humans and animals dynamically interact with their physical and information environments. Although they may have regions or territories that they prefer \cite{dosen2013methodological}, movement in the physical and political sense is a defining characteristic.
	\item \textit{Heading} -- There appears to be a rate-limited alignment component that is needed for a group to coalesce. This is obvious in the physical patters of flocking or schooling, but also manifests in language (e.g. \enquote{political alignment}) \cite{denicola2017understanding} and fashion \cite{curran1999analysis}. 
	\item \textit{Influence} --  Agents within a specified \textit{social influence horizon} (SIH) are capable of influencing each other's orientation and speed in that space, inversely proportional to distance. This interacts with heading, as more aligned agents have more time to influence each other \cite{Olfati-Saber2007}.
\end{enumerate} 

The model produces distinct emergent patterns, by adjusting only the SIH. A zero radius has no interaction, while an infinite radius interacts with the entire population equally. Additionally, these behaviors emerge more easily at lower dimensions. We describe the patterns, shown here as agents on a heatmap, as: 

\begin{figure}[H]
	\centering
	\begin{minipage}{.333\textwidth}
		\centering
		\fbox{\includegraphics[width=12em]{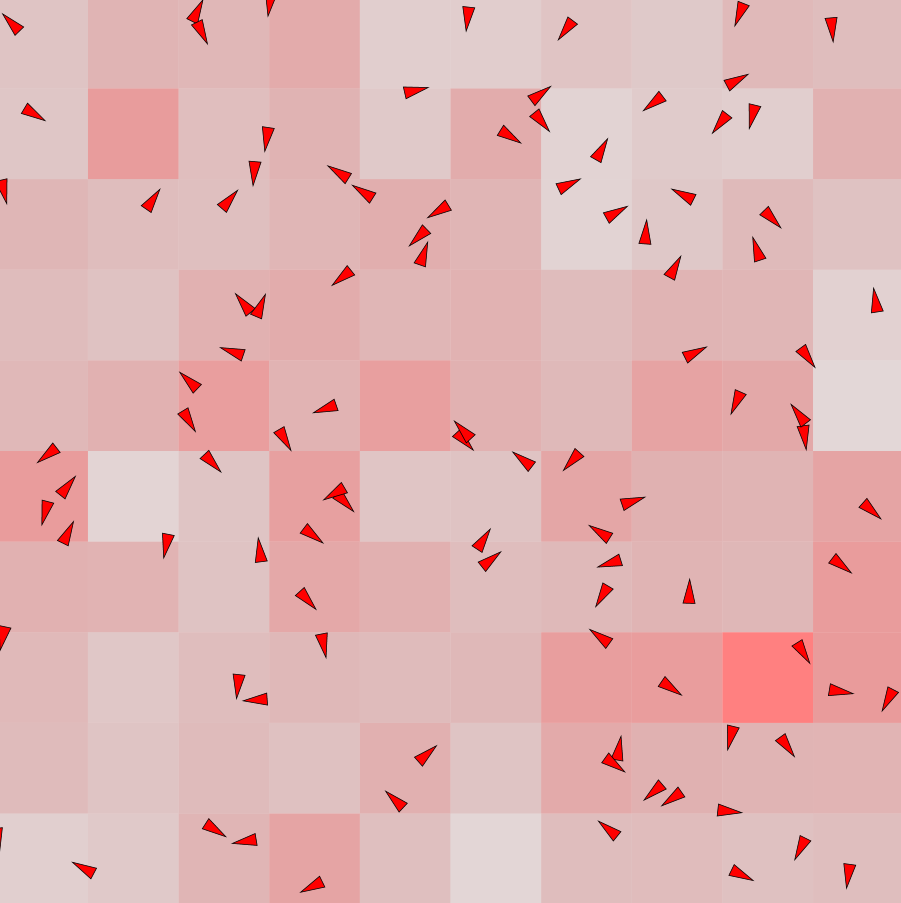}}
		\caption{\label{fig:explorer} Nomadic Phase}
	\end{minipage}%
	\begin{minipage}{.333\textwidth}
		\centering
		\fbox{\includegraphics[width=12em]{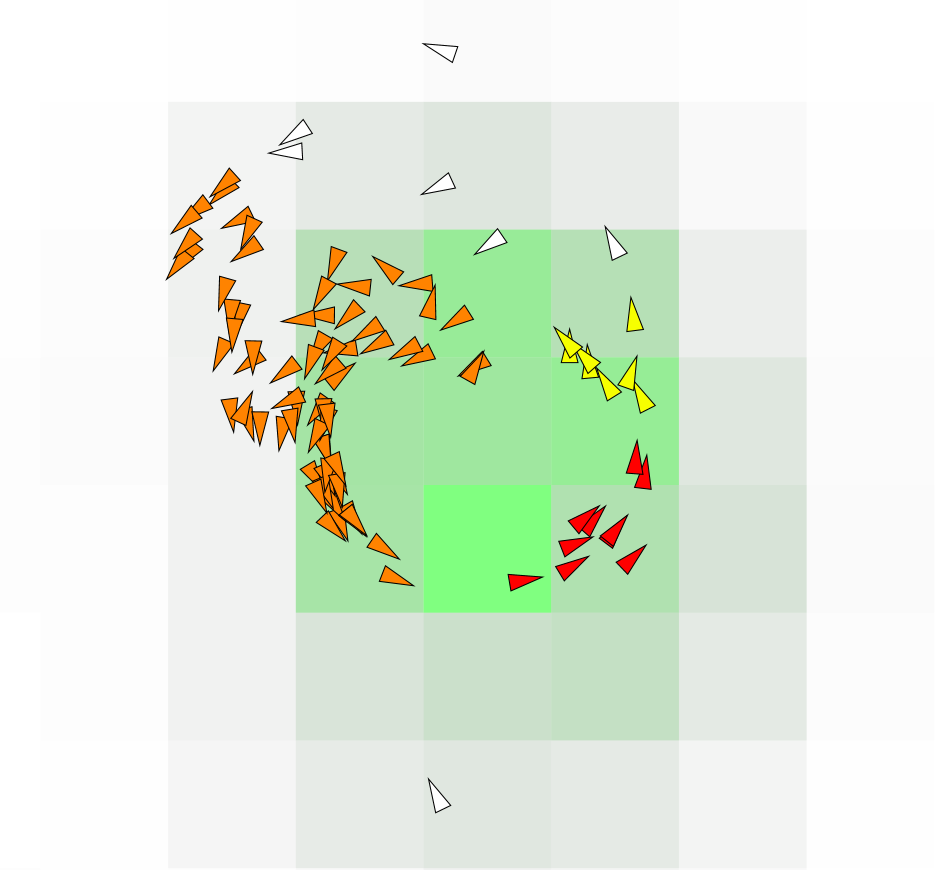}}
		\caption{\label{fig:coloredFlocking} Flocking Phase}
	\end{minipage}%
	\begin{minipage}{.333\textwidth}
		\centering
		\fbox{\includegraphics[width=12em]{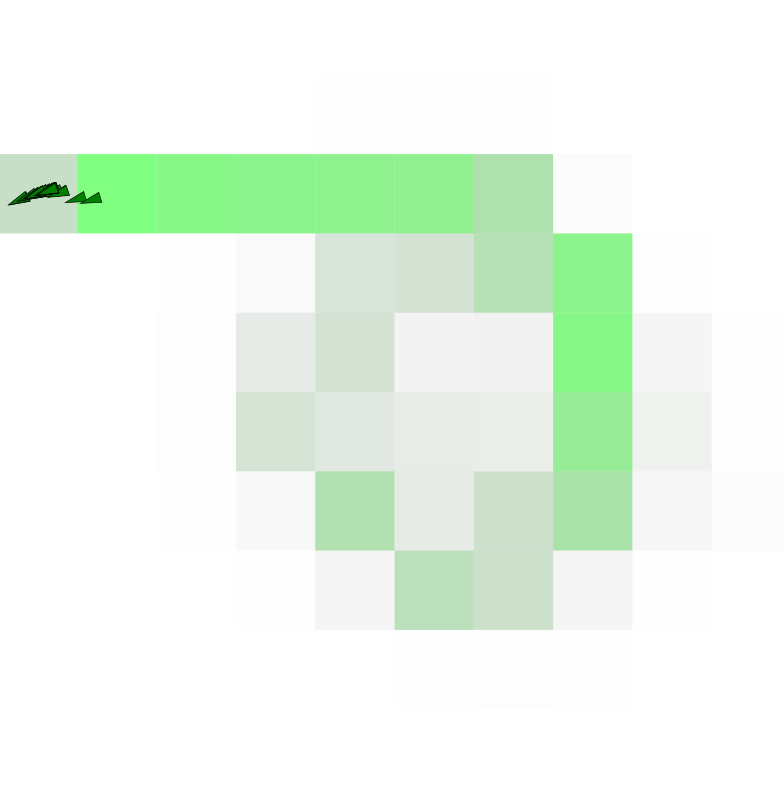}}
		\caption{\label{fig:stampedeScreen} Stampede Phase}
	\end{minipage}%
\end{figure}

\textit{Nomadic Phase} (figure \ref{fig:explorer}) - A low SIH means low influence by other agents, so each agent moves along its own largely independent path.

\textit{Flocking Phase} (figure \ref{fig:coloredFlocking}) - An intermediate SIH results in an agent whose movement is affected by nearby individuals. There is alignment with neighbors, but they do not converge.

\textit{Stampede Phase} (figure \ref{fig:stampedeScreen}) - At high SIH, all members are exposed equally to each other. Alignment converges and supports runaway conditions \cite{lande1981models}. This is represented in other literature as \enquote{filter bubbles}, \enquote{echo chambers} \cite{flaxman2016filter}, \enquote{group polarization}, and \enquote{extremism} \cite{moscovici1994conflict}.

\section{\MakeUppercase{Results}}
\begin{wrapfigure}{R}{0.44\textwidth}
	\centering
	\fbox{\includegraphics[width=20em]{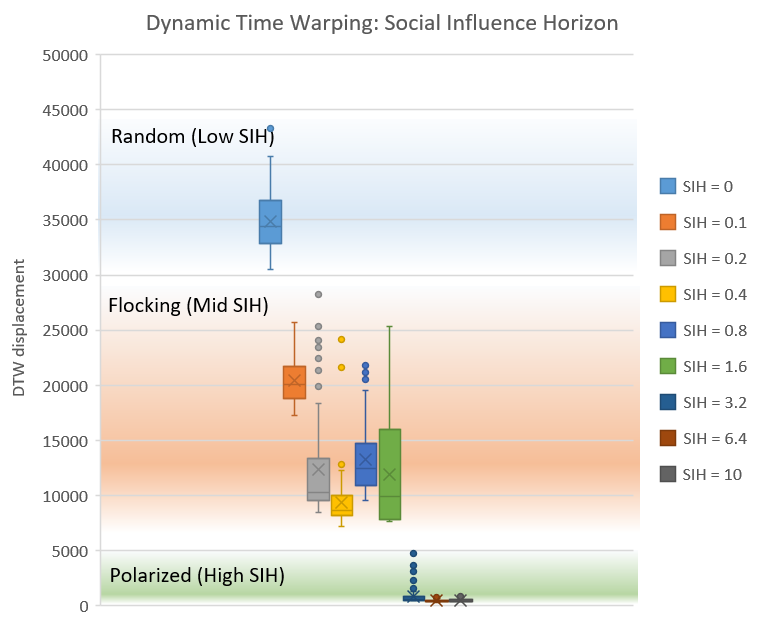}}
	\caption{\label{fig:populations} Nomad, Flock, and Stampede DTW}
\end{wrapfigure}

We used Dynamic Time Warping (DTW) \cite{salvador2007toward} to determine population membership with respect to SIH. DTW attempts to find the lowest distance that one set of points need to be moved to exactly match another sequence of points. The distribution of DTW distance by agent SIH is shown in Figure \ref{fig:populations}. The populations are distinctive and \textit{non-overlapping} in our datasets.

\begin{figure}[h]
	\centering
	\begin{minipage}{.333\textwidth}
		\centering
		\fbox{\includegraphics[width=16.5em]{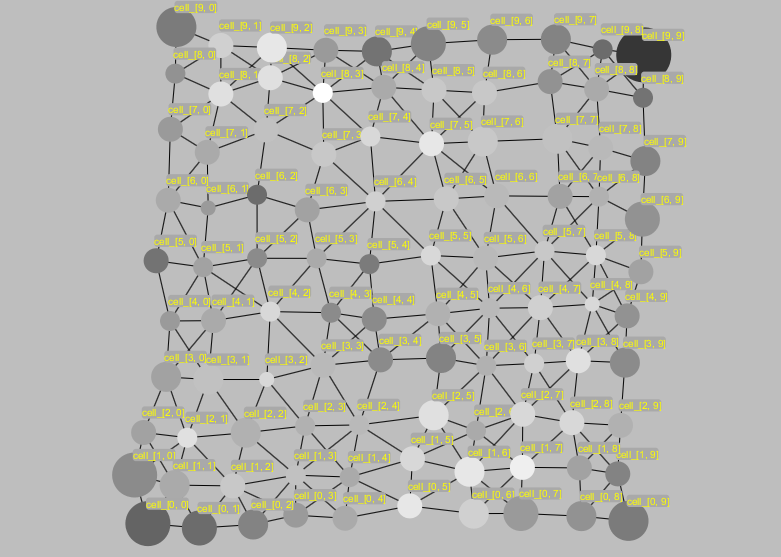}}
		\caption{\label{fig:nomad} Nomad map}
	\end{minipage}%
	\begin{minipage}{.333\textwidth}
		\centering
		\fbox{\includegraphics[width=16.5em]{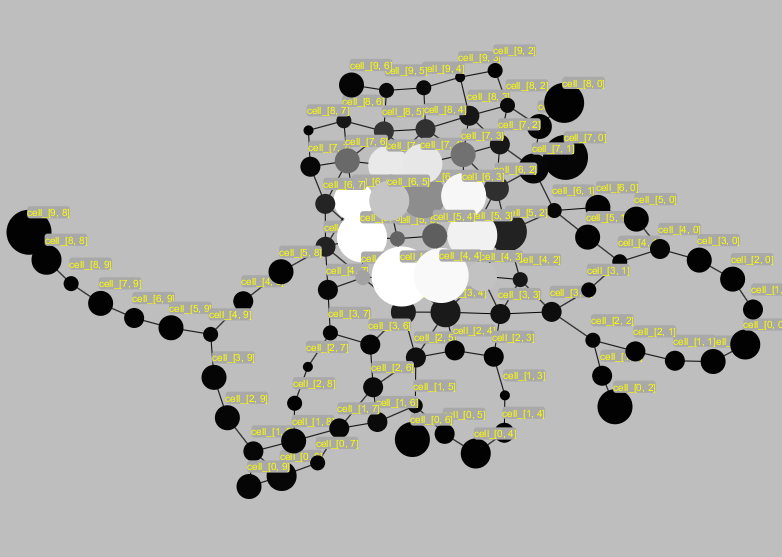}}
		\caption{\label{fig:flocking} Flock map}
	\end{minipage}%
	\begin{minipage}{.333\textwidth}
		\centering
		\fbox{\includegraphics[width=16.5em]{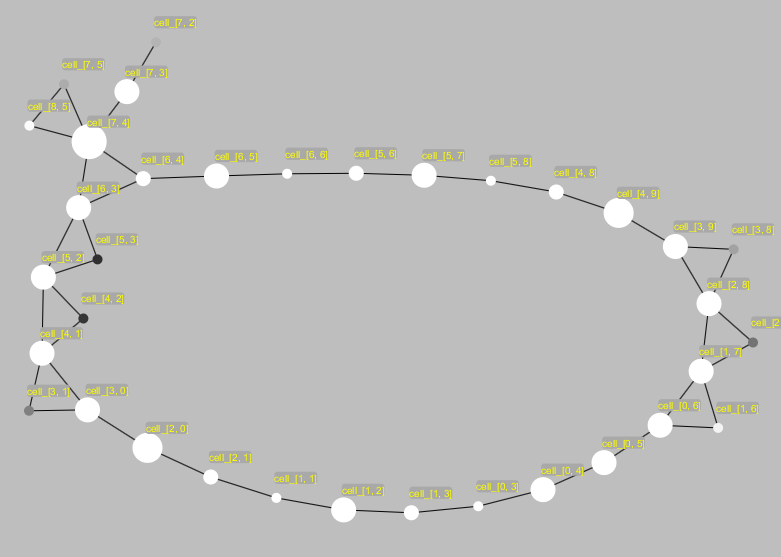}}
		\caption{\label{fig:stampede} Stampede map}
	\end{minipage}%
\end{figure}

The simulator stores a trajectory for each agent that contains the labels of each \enquote{belief cell} that they traversed at each time sample, which serve as a proxy for word embeddings. Like Yao's dynamic Word2Vec work \cite{Yao:2018:DWE:3159652.3159703}, these sets of terms describes the agent's semantic path. To construct maps of the environmental and social features of the belief space, we build a network of term-nodes, using all the agents' trajectories. Each trajectory creates a string of nodes, one for each cell that the agent passed through. Common nodes are shared. As trajectories accumulate, identifiable characteristics emerge. Layout is calculated using the F-R Force-directed algorithm \cite{Fruchterman1991Graph}. Examples of these  constructions are shown in Figures \ref{fig:nomad}, \ref{fig:flocking}, and \ref{fig:stampede}.

In these maps, node size represents the average time that an agent spent \enquote{over} the cell, while brightness represents  unique visitors. These maps reflect the characteristics of their respective populations. The nomadic group reproduces the rectangular shape of the environment. With flocking social influence tends to pull agents towards denser areas away from the borders. This results in a map with detailed popular areas where agents congregate and unexplored areas that do not have a reliable relationship with the underlying terms. The stampeding group has the same environmental awareness of a single agent, but the social inertia of a population. In the map produced by this behavior, the relationship of the trajectories to the underlying coordinate frame is completely lost. 

When we lay out the flocking and stampeding values on the nomadic network, we can see a \textit{belief map} (Fig \ref{fig:anthill}) that shows the bounds of the explored environment and the populations. In the image, cylinder diameter is the average agent dwell time and height is the number of unique visitors. The white nomad population establishes the environment. The flocking population clusters towards the center, reflecting the mix of social and environmental influences. The red cylinders reflect the socially dominated behavior of stampeding agents. The gray mesh is all nomad paths.  

The next goal of this effort will be to validate our theoretical, simulated model against clean, annotated data of computer-mediated human interaction. In success, this should lead to the automated production of maps that allow users to contextualize the beliefs they hold and those they encounter.

\newpage

\bibliographystyle{ci-format}

\end{document}